\documentclass[prb,superscriptaddress,showpacs,preprintnumbers,twocolumn,floatfix]{revtex4}

\usepackage{amsmath}
\usepackage{amssymb}
\usepackage{graphicx}
\newcommand{\nn}{\nonumber}
\newcommand{\kb}{k_{_{\mathrm{B}}}}
\newcommand{\eps}{\varepsilon}
\newcommand{\bp}{\mathbf{p}}
\newcommand{\la}{\left<}
\newcommand{\ra}{\right>}
\newcommand{\Tc}{\ensuremath{T_c}}

\newcommand{\jour}[4]{#1\ \textbf{#2},\ #3\ (#4)}
\newcommand{\Fref}[1]{Fig.~\ref{#1}}
\newcommand{\Eref}[1]{Eq.~(\ref{#1})}
\newcommand{\Rref}[1]{Ref.~\onlinecite{#1}}

\newcommand{\PRL}{Phys. Rev. Lett.}
\newcommand{\PR}{Phys. Rev.}

\begin{document}
\title{Specific Heat Discontinuity in Impure
Two-Band Superconductors}

\author{Todor M. Mishonov}
\email[E-mail: ]{todor.mishonov@fys.kuleuven.ac.be}
\thanks{Phones: +32-16-3-27183, +32-16-224877, Fax: +32-16-327983}
\affiliation{Laboratorium voor Vaste-Stoffysica en Magnetisme,
Katholieke Universiteit Leuven,\\ Celestijnenlaan 200 D, B-3001
 Leuven, Belgium}
\affiliation{Faculty of Physics,
Sofia University ``St. Kliment Ohridski'', 5~J.~Bourchier Blvd.,
1164 Sofia, Bulgaria}
\author{Evgeni S. Penev},
\affiliation{Faculty of Physics,
Sofia University ``St. Kliment Ohridski'', 5~J.~Bourchier Blvd.,
1164 Sofia, Bulgaria}
\author{Joseph O. Indekeu}
\email[E-mail: ]{joseph.indekeu@fys.kuleuven.ac.be}
\thanks{Phone: +32-16-327127}
\affiliation{Laboratorium voor Vaste-Stoffysica en Magnetisme,
Katholieke Universiteit Leuven,\\ Celestijnenlaan 200 D, B-3001
 Leuven, Belgium}
\author{Valery L. Pokrovsky}
\email[E-mail: ]{valery@physics.tamu.edu}
\affiliation{Department of Physics, Texas A\&M
University, College Station, Texas 77843-4242\\ and Landau Institute
for Theoretical Physics, Kosygin Street~2, Russia, Moscow~117940}

\date{\today}
\begin{abstract}
The Ginzburg-Landau coefficients, and the jump of the specific
heat are calculated for a disordered two-band superconductor. We
start with the analysis of a more general case arbitrary
anisotropy. While the specific heat discontinuity at the critical
temperature $T_c$ decreases with increasing disorder, its ratio to
the normal state specific heat at \Tc\ increases and slowly
converges to the isotropic value. For a strong disorder the
deviation from the isotropic value is proportional to the elastic
electron scattering time. In the case of a two-band superconductor
we apply a simplified model of the interaction independent on
momentum within a band. In the framework of this model all
thermodynamic values can be found explicitly at any value of the
scattering rate. This solution explains the sample dependence of
the specific heat discontinuity in MgB$_2$ and the influence of
the disorder on the critical temperature.
\end{abstract}


\maketitle

\section{Introduction}
\label{sec:1}

The investigation of the specific heat $C(T)$ is an important tool
for understanding the nature of the superconductivity and
anisotropy of the superconducting gap $\Delta_\bp(T)$ on the Fermi
surface $\eps_{\bp} = E_{\mathrm{F}}$. Historically the relative
specific heat jump $\Delta C/C_n(\Tc)$ was used to establish the
BCS picture\cite{bcs:57} for the conventional superconductors
having nearly isotropic gap. Subsequently the thermodynamics of
clean anisotropic-gap superconductors was analyzed in the weak
coupling approximation by Pokrovsky and Ryvkin.
\cite{Pokrovsky:61,Pokrovsky:63} They have found that anisotropy
suppresses the value $\Delta C/C_n(\Tc)$ in comparison to its
isotropic value 1.43. This inequality is not satisfied in
classical low-temperature superconductors partly because they are
not extremely clean, but also since the weak coupling
approximation has a poor precision. Geilikman and Kresin
\cite{Geilikman:61} have proved that the first correction due to
interaction increases $\Delta C/C_n(\Tc)$ and thus disguises the
anisotropy effect. The modern superconductors display really high
anisotropy. In particular, superconductivity is highly anisotropic
in MgB$_2$. This fact is the main motivation of this work. It is
well known that only superconducting crystals of very high quality
can reach the theoretical clean-limit asymptotics. As a rule, the
reduced specific heat jump is sample-dependent and understanding
of this disorder dependence is a challenging problem. The latter
is especially important for MgB$_2,$ a compound now being in the
limelight of superconductor materials science.

The aim of the present paper is to derive the dependence of the
critical temperature and the relative specific heat jump $\Delta
C/C_n(\Tc)$ on the elastic scattering time of the charge carriers
at the critical temperature $\tau(\Tc),$ for two-band
superconductors having in mind application to MgB$_2$. For this
purpose we need corresponding formulae for a general dirty
anisotropic superconductor. Such equations were derived in
\cite{Pokrovsky:96}. We reproduce them here for readers
convenience and because there occurred several misprints in the
cited work which we correct here. In the $\tau$-approximation the
electrical resistivity of the normal metal $\rho_{\mathrm{el}}$ is
determined by this scattering time. Thus, our formula can be used
for the investigation of correlations in the experimentally
determined $\Delta C/C_n(\Tc)$ versus
$\rho_{\mathrm{el}}(\Tc)/\Tc$ plot. The comparison of the
theoretical curve and the experimental data can reveal the gap
anisotropy $\Delta_{b,\bp}$ and the scattering rate $1/\tau(\Tc)$.
The gap can depend on both the quasimomentum $\bp$ and the band
index $b.$

The applicability of the weak-coupling theory to MgB$_2$ is
contentious. However, experimental results on the relative specific
heat discontinuity\cite{Wang:01} indicate that the anisotropy effect
is more profound than the effect of interaction. For this compound,
the reduced specific heat $\Delta C/C_n(\Tc)$ is definitely smaller
than the weak-coupling BCS value 1.43. Moreover the temperature
dependence of the specific heat of the superconducting phase is
described fairly well\cite{Bouquet:01} by the two-band
model\cite{Moskalenko:59} and the relative specific heat jump
agrees\cite{Bouquet:01} with the Moskalenko's weak-coupling
formula.\cite{Moskalenko:59} The comparison of the latter and the
\textit{ab initio} strong-coupling calculations\cite{Golubov:02} for
MgB$_2$ shows that the decrease of $\Delta C/C_n(\Tc)$ due to
different values of the superconducting gap for different bands is
at least 2 times bigger than the increase of this reduced specific
heat jump due to the strong coupling effects.  We discuss this point
in the concluding discussion.

\section{Clean superconductors}
\label{sec:2}

In this section we reproduce some results for an anisotropic clean
superconductors obtained by different authors many years ago and
derive equation for the specific jump discontinuity in this case.
Though neither of these results is new, they are necessary for
understanding nest sections.It was shown in \Rref{Pokrovsky:63}
that, within the framework of the weak-coupling theory, the order
parameter possesses the property of separability:
\begin{equation}
  \Delta_{\bp}(T) = \Xi(T)\chi_{\bp}.
\label{sep}
\end{equation}
According to \Eref{sep}, the temperature dependence characterized by
the factor $\Xi(T)$ is separated from the angular dependence described
by the factor $\chi_\bp$. The Ginzburg-Landau (GL) expansion for the
free energy density\cite{LLV,GL,LL:9} can be written in terms of the
temperature-dependent factor $\Xi(T)$ alone:
\begin{equation}
f(\Xi,T)=a_0\frac{T-\Tc}{\Tc}\,|\Xi|^2+\frac{1}{2}b|\Xi|^4.
\end{equation}
The specific heat jump per unit volume is related to the GL
coefficients by the following relation:
\begin{equation}
 (C_s-C_n)|_{\Tc} = \Delta C = \frac{1}{\Tc} \frac{a_0^2}{b},
\label{eq:DeltaC}
\end{equation}
where $C_s$ is the specific heat per unit volume of the
superconducting phase and $C_n$ is that of the normal phase.

Our starting point are the expressions of Gor'kov and
Melik-Barkhudarov\cite{Gorkov:64} for the GL coefficients in the clean
limit which can be written as
\begin{equation}
\label{eq:GLclean}
 a_0 = \nu_{\mathrm{F}} \la{\chi^2}\ra,\quad
 b = \frac{\zeta(3,\frac{1}{2})
     \nu_\mathrm{F}}{2(2\pi\kb\Tc)^2}\,\la{\chi^4}\ra,
\end{equation}
where the Hurwitz and the Riemann zeta functions, $\zeta(k,z)$ and
$\zeta(k)$, respectively, read
\begin{equation}
\zeta(k,z)=\sum_{n=0}^{\infty}(n+z)^{-k},\quad
\zeta(k)=\zeta(k,1)=\sum_{n=1}^{\infty}n^{-k},
\end{equation}
and obey the relation
$\zeta\left(k,\frac{1}{2}\right)=(2^k-1)\zeta(k).$ A simple
variational derivation of \Eref{eq:GLclean} is given in
\Rref{Mishonov:02}.  The celebrity of $\zeta(3)$ in mathematics has
been discussed in \Rref{Borwein:00}.  The normalized moments of the
gap anisotropy function are determined by averaging over the Fermi
surface, having the general form in the $D$-dimensional case
\begin{align}
\la{\chi^n}\ra
  &\equiv \idotsint\limits_{\mathrm{BZ}}
   \chi^n_{\bp}\,\delta(\eps_{\bp}-E_{\mathrm{F}})
   \frac{d\bp}{\nu_{\mathrm{F}}(2\pi\hbar)^D} \nn \\
  &=\idotsint\limits_{\eps_{\bp} = E_{\mathrm{F}}} \chi^n_\bp\ \frac{d
 S_{\bp}}{\nu_{\mathrm{F}} v_{\bp} (2\pi\hbar)^D},
\label{eq:mean}
\end{align}
where $dS_{\bp}$ is an infinitesimal surface element and
$\mathbf{v}_{\bp}=\nabla_{\bp}\eps_{\bp}$ is the quasiparticle
velocity. The quasi-momentum space integral is taken over the whole
Brillouin zone (BZ). The integration over the Fermi surface
$\eps_{\bp} = E_{\mathrm{F}}$ implicitly includes summation over
fragments and sheets of different bands, if any. The normalizing
factor
\begin{equation}
\nu_{\mathrm{F}} =\idotsint\limits_{\eps_{\bp} = E_{\mathrm{F}}} \frac{d
 S_{\bp}}{v_{\bp}(2\pi\hbar)^D},
\label{eq:DOS}
\end{equation}
is the density of states (DOS) per unit volume for fixed spin, and
enters the normal-phase specific heat
\begin{equation}
 C_n(T) =\frac{2}{3}\pi^2\kb^2 \nu_{\mathrm{F}}T.
\end{equation}
This equation together with the formulae for the GL coefficients,
\Eref{eq:GLclean}, lead to the following value for the reduced jump of
the specific heat
\begin{equation}
 \left.\frac{\Delta C}{C_n}\right|_{\Tc}
 = \frac{12}{7\zeta(3)} \frac{1}{\beta_{\Delta}},\quad
\begin{cases}\displaystyle
   \frac{1}{\beta_{\Delta}}  =
   \frac{\la{\Delta_\bp^2}\ra^{\,2}} {\la\Delta_{\bp}^4\ra}
   =\frac{\la{\chi^2}\ra^{\,2}}{\la{\chi^4}\ra} \leqslant 1,\\
    \displaystyle
   \frac{12}{7\zeta(3)} = 1.42613\dotsc,
\end{cases}
\label{eq:jump}
\end{equation}
which is exactly the result obtained in
Refs.~\onlinecite{Pokrovsky:61,Pokrovsky:63}; for a methodical
derivation see \Rref{Mishonov:02}. Using \Eref{eq:GLclean}
for $T$ slightly lower than $\Tc,$ we get for the equilibrium order
parameter
\begin{gather}
\label{eq:asdf}
|\Xi|_{\mathrm{eq}}^2 = -\frac{T-\Tc}{\Tc}\frac{a_0}{b},\nn\\
|\Delta_\bp|^2        = |\Xi|_{\mathrm{eq}}^2\chi_\bp^2
    =\frac{2(2\pi\kb\Tc)^2}{\zeta(3,\frac{1}{2})}\frac{\Tc-T}{\Tc}
    \,\frac{\la{\chi^2}\ra}{\la{\chi^4}\ra}\chi_{\bp}^2,
\end{gather}
which is the result by Gor'kov and Melik-Barkhudarov.\cite{Gorkov:64}

\section{Disordered anisotropic superconductors}
\label{sec:3}

\subsection{Transition line and order parameter}

In this subsection we analyze the transition temperature \Tc\ as a
function of the elastic scattering rate $1/\tau$ and the angular
dependence of the order parameter $\chi_{\bp}.$ As it was
explained before, the angular dependence is the same for any
temperature at fixed $\tau$. The transition line has been studied
in \Rref{Pokrovsky:96}.  Although the equations obtained in the
latter work were rather general, their treatment was focused on a
specific situation---a mixture of $s$- and $d$-pairing
characteristic for cuprate superconductors. Therefore, it is
useful to analyze the results for a less exotic case of
anisotropic $s$-pairing. The general equation for the transition
line found in \Rref{Pokrovsky:96} reads
\begin{equation}
 g(\Tc,\tau )\sum_n \frac{V_n |\la{\Psi_n}\ra|^2
  }{1-f(\Tc,\tau)V_n} =1,
\label{T-line}
\end{equation}
where following notations are introduced:
\begin{align}
f(T,\tau) &= \frac{1}{\pi}\left[\ln\frac{\bar{\epsilon}}
    {2\pi\kb T}-\digamma\!\left(x+\frac{1}{2}\right)\right],\label{f}\\
g(T,\tau) &= \frac{1}{\pi}\left[\digamma\!\left(x+\frac{1}{2}\right)
    -\digamma\!\left(\frac{1}{2}\right)\right],
\label{g}
\end{align}
$x=(2\pi \kb T\tau/\hbar)^{-1}$; $\bar{\epsilon}$ is the cutoff
energy; $\digamma(x)$ is the Euler digamma function; $V_n$ are
eigenvalues of the linear operator $\hat{V}$ with kernel
$V(\bp,\bp^{\prime})$ equal to the electron-electron effective
interaction energy at the Fermi surface multiplied by the DOS
$\nu_\mathrm{F}$; $\Psi_n({\bp})$ are the corresponding eigenfunctions
normalized according to the condition $\la{|\Psi_n|^2}\ra=1$. The
transition temperature of the clean superconductor is determined by
the equation $f(T_{c0},\tau=\infty)=V_0^{-1}$ which gives
\begin{equation}
\kb T_{c0}=
\frac{2\gamma\bar{\epsilon}}{\pi}\exp\left(-\pi /V_0\right),
\qquad \chi\equiv\Psi_0,
\label{clean}
\end{equation}
where
\begin{equation}
\gamma=\exp{\mathrm{C}}
=\frac{1}{4\exp\left(\digamma(1/2)\right)}
=1.781, 
\end{equation}
$\mathrm{C}=0.577$ 
is the Euler constant,
and $V_0$ is the maximum eigenvalue of the operator $\hat{V}.$ The
angular dependence of the order parameter in this case is given by the
corresponding eigenfunction $\Psi_0({\bp})$. As long as $x_0=(2\pi\kb
T_{c0}\tau/\hbar)^{-1}$ remains large the transition temperature \Tc\
differs from $T_{c0}$ by insignificant corrections of the order of
$x_0^{-1}.$ We call the superconductor moderately dirty if the value
$x_0$ becomes small, but $f(\Tc,\tau)$ is still close to
$V_0^{-1}$. More precisely, it means that
$|\ln\bar{\epsilon}\tau/\hbar-\pi V_0^{-1}| \ll V_0^{-1}$.
Then the solution of \Eref{T-line} reads
\begin{equation}
\Tc(\tau) = T_{c0}\left(\frac{\tau}{\tau_0}\right)^{\kappa -1},
\label{hohenberg}
\end{equation}
where $\kappa=\la{\chi^2}\ra/\la{\chi}\ra^2 \geq 1$ is the anisotropy
coefficient, which is equal to 1 for an isotropic superconductor, and
$\tau_0=2\hbar\gamma/(\pi\kb T_{c0}).$ In this range of the scattering
rate the angular dependence of the gap $\chi_\bp$ is identical to that
for the clean superconductor: $\chi_\bp=\Psi_0({\bp})$.  Thus, the
transition temperature decreases in a power-like way with the increase
of the scattering rate $1/\tau$ or the residual resistivity
$\rho_\mathrm{res}$ proportional to this rate. This is a peculiarity
of the anisotropic superconductor. The exponent in \Eref{hohenberg} is
zero for the isotropic superconductor,
cf. \Rref{Anderson:59}. Equation~(\ref{hohenberg}) was first derived
by Hohenberg\cite{Hohenberg:63} for weakly anisotropic
superconductors. Its validity for arbitrary $\kappa$ in the range of
moderate dirt was proved in \Rref{Pokrovsky:96}.

We call the dirt strong if the parameter
$\ln\bar{\epsilon}\tau/\pi\hbar$ becomes less than $V_0^{-1}$ and has
the order of magnitude of $V_0^{-1},$ and if the difference
$V_0^{-1}-\ln\bar{\epsilon}\tau /\pi\hbar$ is not small in comparison
to $V_0^{-1}$. Equation (\ref{hohenberg}) remains qualitatively
correct, but $\kappa$ becomes a slowly varying function of $\tau$. The
exact formula for the transition temperature in this range is given by
Eq.~(37) of \Rref{Pokrovsky:96}.

In the extra-dirty limit $\frac{1}{\pi}\ln\bar{\epsilon}\tau/\hbar $
becomes much smaller than $V_0^{-1}$, but still
$\bar{\epsilon}\tau/\hbar\gg 1.$ The last inequality ensures that the
elastic scattering does not destroy the Fermi surface. In the
extra-dirty limit the angular dependence of the gap reaches its
limiting value $\chi_\bp\propto\bar{V}({\bp})$, where
$\bar{V}({\bp})=\la V(\bp,\bp^{\prime})\ra_{{\bp}^{\prime}}$.  The
equation for \Tc\ in the extra-dirty limit reads:
\begin{equation}
\kb \Tc (\tau) = \frac{2\gamma\bar{\epsilon}}{\pi}
    \exp\left(-\frac{\pi}{\langle\bar{V}\rangle}\right)
    (\bar{\epsilon}\tau/\hbar)^{\bar{\kappa} -1},
\label{extra}
\end{equation}
where $\langle\bar{V}\rangle=\la\bar{V}({\bp})\ra_\bp.$ It should be
noted that $\bar{\kappa}$ in the last equation differs from that for
the clean superconductor, namely:
\begin{equation}
\bar{\kappa}=\frac{\sum_n V_n^2 \la{\Psi_n}\ra^2}{\left(\sum_n
V_n\la{\Psi_n}\ra\right)^2}.
\label{kappa}
\end{equation}

\subsection{Specific heat discontinuity}

The theory of dirty anisotropic superconductors
\cite{Pokrovsky:96} was based on the Green's functions method
combined with the Abrikosov-Gor'kov averaging over the random
impurity field.\cite{AG,AGD} A simplifying assumption was the
isotropy of the scattering which is characterized by a constant
rate $1/\tau$. In particular, the authors derived the GL equations
and GL coefficients with an accuracy of a common scaling
factor\cite{Pokrovsky:96} (see also an earlier article
\cite{Muzikar} For the representation accepted in this article
this factor is $\la{\chi^2}\ra$, as it follows from the comparison
of \Eref{eq:GLclean} and Eqs.~(48), (60), (78)--(82) in
\Rref{Pokrovsky:96}. Correcting a misprint in Eq.~(59) in
\Rref{Pokrovsky:96}, further repeated in Eqs.~(61), (82) therein,
and slightly regrouping terms we find:
\begin{equation}
 a_0 = \nu_{\mathrm{F}}
 \left[\la{\chi^2}\ra-\left(\la{\chi^2}\ra-\la{\chi}\ra^{2}
 \right)x_c\zeta_{2,0}\right],
\label{a}
\end{equation}
\begin{eqnarray}
b&=& \frac{\nu_{\mathrm{F}}}{8(\pi k_\mathrm{B}T_{\mathrm{c}})}
\left[\left\langle \chi ^{4}\right\rangle \zeta_{3,0}
-\left\langle \chi^{2}\right\rangle^{2}x_{c}\zeta_{4,0}\right.
\nonumber\\
&&\qquad\qquad\quad
+4\left\langle \chi^{3}\right\rangle
  \left\langle \chi \right\rangle x_{c}\zeta_{3,1}\nonumber\\[5pt]
&&\qquad\qquad\quad
+2\left\langle \chi^{2}\right\rangle \left\langle \chi \right\rangle^{2}
  \left(x_{c}^{3}\zeta_{4,2}+x_{c}^{2}\zeta_{3,2}\right)\nonumber\\[5pt]
&&\qquad\qquad\quad\left.
+\left\langle \chi \right\rangle^{4}x_{c}^{4}\zeta_{4,3}\right],
\label{eq:dirty}
\end{eqnarray}
where
\begin{equation}
x_c=\frac{\hbar/\tau(\Tc)}{2\pi\kb\Tc} = \frac{x_0}{\Tc/T_{c0}},
\end{equation}
is the dimensionless scattering rate extrapolated to the critical
temperature. The resistivity of the normal metal $\rho_{\text{el}}$ is determined by the Drude formula:
\begin{equation}
\rho_{\text{el}}^{-1}=m^{-1}n_\mathrm{tot}e^2\tau,
\end{equation}
where $m$ is the effective mass tensor and $n_\mathrm{tot}$ is the
density of normal charge carriers; for clean crystals the total
volume density of all charge carriers, electrons and holes,
$en_\mathrm{tot}=e(n_e-n_h)$, can be determined by the Hall constant
${\cal R}_{\mathrm{H}}=1/en_\mathrm{tot}$ in strong magnetic fields.
For clean superconductors, disregarding some subtleties, the same
ratio $n/m$ enters the London penetration depth
$\lambda_{\text{clean}}$ at $T=0$
\begin{equation}
\frac{1}{\lambda_{\text{clean}}^2(0)}=\frac{ne^2}{mc^2\eps_0},
\end{equation}
where in Gaussian units $\eps_0=1/4\pi;$ $\lambda_{\text{clean}}(0)
\sim$~0.1--1~$\mu$m. Multiplying these equations we obtain a useful
estimate
\begin{equation}
 x_c\simeq\frac{\hbar
 c^2\eps_0\rho_{\text{el}}(\Tc)}{2\pi\lambda_{\text{clean}}^2(0)\kb\Tc}.
\label{eq:xEvaluation}
\end{equation}
The notation $\zeta_{k,l}$ in \Eref{eq:dirty} stands for the
generalized zeta-functions defined in \Rref{Pokrovsky:96} and taken at
the value of the argument $z_c=x_c+1/2$, i.e.\cite{note}
\begin{equation}
 \zeta_{k,l} \equiv \zeta_{k,l}(z_c).
\end{equation}
For the readers' convenience we recall the definition of these
functions:
\begin{equation}
\zeta_{k,l}(z)=\sum_{n=0}^{\infty}(n+z)^{-k}(n+1/2)^{-l}.
\label{eq:zkl}
\end{equation}
They represent a natural generalization of the Hurwitz zeta functions:
\begin{equation}
\zeta_{k,0}(z)=\zeta(k,z),\quad \zeta_{k,0}(1)=\zeta(k,1)=\zeta(k).
\end{equation}
Below we provide the asymptotics of $\zeta_{k,l}(z)$ for
$z\rightarrow\infty$ necessary for the further calculations:
\begin{alignat}{3}
\zeta_{k,l}(z) \sim &(2^l-1)\zeta(l)z^{-k}
-k[(2^{l-1}-1)\zeta(l-1)\nn\\&
\qquad-(2^l-1)\zeta(l)/2]z^{-k-1}, \quad
&\text{for}\quad & l\ge2,\nn\\ \zeta_{k,1}(z) \sim & z^{-k}\ln z,
\quad &\text{for}\quad & k\ge1,\nn\\ \zeta_{k,0}(z) \sim &
(k-1)^{-1}z^{-k+1},   \quad &\text{for}\quad & k  >1.
\label{eq:ZetaAsymp}
\end{alignat}
Let us note that, for integer arguments $k$, the Hurwitz
zeta functions are associated with the Euler polygamma function
$\digamma^{(k)}$:
\begin{equation}
\zeta(k+1,z)=\frac{(-1)^{k+1}}{k!}\digamma^{(k)}(z),\quad k=1,2,3,\dots.
\end{equation}

With these notations the reduced discontinuity of the specific heat reads:
\begin{align}
\left.\frac{\Delta C}{C_n}\right|_{\Tc}
 = \frac{a_0^2}{\kb\Tc b}
 = \frac{12}{7\zeta(3)}\frac{1}{\beta_\tau},
\label{eq:jumpdirty}
\end{align}
where
\begin{align}
\label{eq:beta}
 \frac{1}{\beta_\tau} = & 7\zeta(3)
\left[\la{\chi^2}\ra-\left(\la{\chi^2}\ra-\la{\chi}\ra^{2}
 \right)x_c\zeta_{2,0}\right]^2\nn\\[5pt]
&\quad\;\times \left[
\left\langle \chi ^{4}\right\rangle \zeta_{3,0}
-\left\langle \chi^{2}\right\rangle^{2}x_{c}\zeta_{4,0}\right.
\nn\\[5pt]
& \qquad\quad
+4\left\langle \chi^{3}\right\rangle
  \left\langle \chi \right\rangle x_{c}\zeta_{3,1}\nn\\[5pt]
& \qquad\quad
+2\left\langle \chi^{2}\right\rangle \left\langle \chi \right\rangle^{2}
  \left(x_{c}^{3}\zeta_{4,2}+x_{c}^{2}\zeta_{3,2}\right)\nonumber\\[5pt]
& \qquad\quad
\left.
+\left\langle \chi \right\rangle^{4}x_{c}^{4}\zeta_{4,3}
   \right]
   ^{-1}.
\end{align}
This general equation will be applied in the following subsections to
some important special cases.

It should be stressed that for isotropic superconductors,
$\la\chi^n\ra=1$, the specific heat jump is impurity independent:
$\beta_\tau=1$. The proof is straightforward taking into account the
identity
\begin{align}&
\zeta_{3,0}
-x_{c}\zeta_{4,0}
+4x_{c}\zeta_{3,1}
+2\left(x_{c}^{3}\zeta_{4,2}+x_{c}^{2}\zeta_{3,2}\right)
+x_{c}^{4}\zeta_{4,3}\nn\\
&=\zeta(3,1/2)=7\zeta(3).
\end{align}
Likewise, using \Eref{eq:ZetaAsymp} one can prove that the asymptotic
form of $\Delta C/C_n(\Tc)$ for an extremely disordered superconductor
with an arbitrary anisotropy is given by, to leading order in
$x_c^{-1}$,
\begin{equation}
\frac{\Delta C}{C_n(\Tc)} \sim\frac{12}{7\zeta(3)}\left [
1-\frac{(\kappa -1)}{x_c}\left (\frac{5\pi^2}{14\zeta(3)}-1\right
)\right ].
\end{equation}

\subsection{Two-band superconductors}

\subsubsection{Critical curve and order parameter}

Keeping in mind the application to MgB$_2$ (for a review see
\Rref{Buzea:01}), we apply the general results of the previous
sections to a simplified model of a two-band superconductor. In this
model we assume that the Fermi surface consists of two disconnected
sheets having different DOS.  The interaction amplitude $V(\bp,\bp')$
is assumed to be a constant within each band. Thus, it can be
described by a $2\times2$ matrix:
\begin{equation}
\hat{V}=\left({W_1\atop U}{U\atop W_2}
\right),
\label{2b-int}
\end{equation}
where $W_1,W_2$ are the interaction energies between any two points
within the first and the second sheet of the Fermi surface,
respectively; $U$ is the interaction between any two points of
different bands.

\begin{figure}[t]
\centering
\includegraphics[width=0.8\columnwidth]{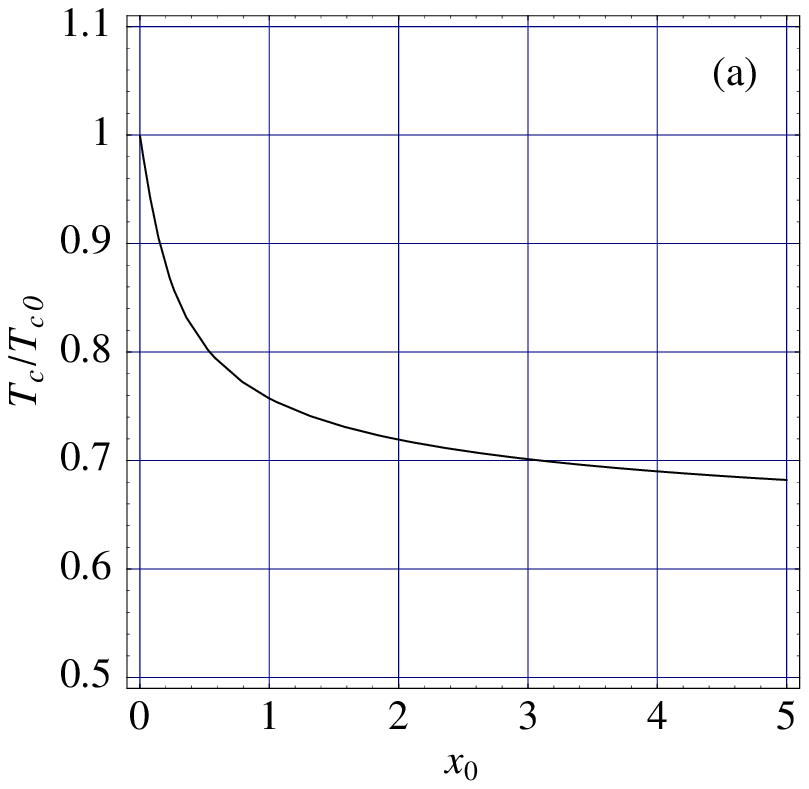}\\
\includegraphics[width=0.8\columnwidth]{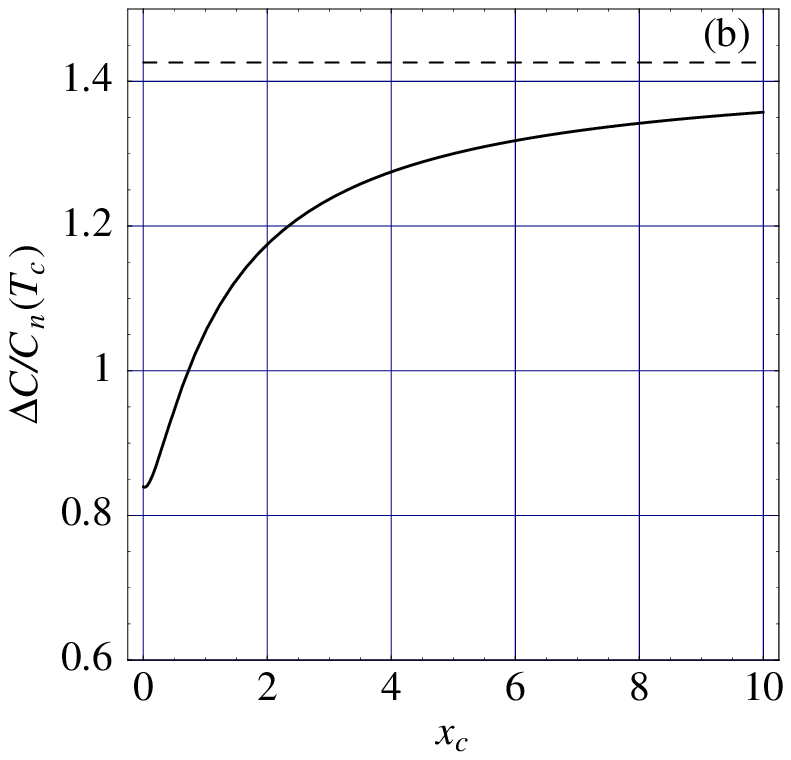}
\caption{(a) Critical curve $\Tc(\tau)$ for two-band model: the
reduced critical temperature $\theta=\Tc(\tau)/T_{c0}$ versus
dimensionless scattering rate $x_0=\hbar/2\pi\kb T_{c0}\tau\propto
\rho_\mathrm{el}(\Tc)$. The set of parameters corresponds to MgB$_2$;
for details see the text.
(b) Reduced specific heat jump $\Delta C/C_n(\Tc)$ as a function of
the dimensionless scattering rate $x_c = \hbar/2\pi\kb\Tc\tau\propto
\rho_\mathrm{el}(\Tc)/\Tc$ for the same set of parameters. The dashed
line indicates the asymptotic (BCS) value for $x_c \gg 1$. The curve
will be shifted up by about 20\% due to the strong-coupling effects.
\label{fig:1}}
\end{figure}

Let us first work out the transition temperature $T_{c0}$ and the
order parameter $\chi$ for a clean two-band
superconductor.\cite{Moskalenko:59} For our simplified model the order
parameter $\chi (\bp)$ is a constant within each band, i.e. it can be
represented by a 2-component vector:
$$
\chi =\left({\chi_1 \atop \chi_2} \right)
$$
The eigenvectors $\Psi$ of the operator $\hat{V}$ obey the following
linear equations:
\begin{align}
c_1W_1\Psi_1 + c_2U\Psi_2   &=\lambda\Psi_1,\nonumber\\
c_1U\Psi_1   + c_2W_2\Psi_2 &=\lambda\Psi_2, \label{2b-eqs}
\end{align}
where the coefficients $c_{1,2}=\nu_{1,2}/(\nu_1+\nu_2)$ are the
statistical weights of the two bands, which reflect the integral
character of the operator $\hat{V}$. The two independent eigenvalues
of Eqs.~(\ref{2b-eqs}) read
\begin{equation}
V_{0,1}\equiv\lambda_{\pm}=\eta\pm\epsilon,
\label{2b-eigenvalues}
\end{equation}
where
\begin{gather}
\eta =\frac{1}{2}(c_1W_1+c_2W_2),\quad
\epsilon=\sqrt{\xi^2+c_1c_2U^2},\nn\\
\xi =\frac{1}{2}(c_1W_1-c_2W_2).
\end{gather}
The corresponding eigenvectors are:
\begin{align}
\Psi_+ &= \left({\sqrt{\frac{1}{2c_1} \left(1 + \xi/\epsilon\right)}}
 \atop
\mathrm{sign}(U) \sqrt{\frac{1}{2c_2}\left(1 - \xi/\epsilon\right)}\right),
\label{Psi+}\\
\Psi_- & = \left({ -\mathrm{sign}(U)
 \sqrt{\frac{1}{2c_1}\left(1- \xi/\epsilon\right)}\atop
 \sqrt{\frac{1}{2c_2}\left(1+ \xi/\epsilon\right)}}\right). \nn
\end{align}
This is apparently a kind of the Bogolyubov transformation. Both
vectors are normalized:
\begin{equation}
\langle\left|\Psi_{\pm}\right|^2\rangle
=c_1\left|\Psi_{\pm1}\right|^2+c_2\left|\Psi_{\pm2}\right|^2=1.
\end{equation}
The average values of the anisotropic eigenfunctions read:
\begin{align}
 \langle\Psi_+\rangle &=
 \sqrt{\frac{c_1}{2}\left(1+\frac{\xi}{\epsilon}\right)}+
 \mathrm{sign}(U)\sqrt{\frac{c_2}{2}\left(1-\frac{\xi}{\epsilon}\right)},
 \label{av+}\\
\langle\Psi_-\rangle &=
 -\mathrm{sign}(U)\sqrt{\frac{c_1}{2}\left(1-\frac{\xi}{\epsilon}\right)}+
 \sqrt{\frac{c_2}{2}\left(1+\frac{\xi}{\epsilon}\right)}.
\label{av-}
\end{align}
It is useful to write simple expressions for the squared averages:
\begin{align}
\langle\Psi_{\pm}\rangle^2&=\frac{1}{2}\pm\frac{1}{2\epsilon}[(c_1-c_2)\xi
+2c_1c_2U],\\ \
\la\Psi_{+}\ra^2&+\la\Psi_{-}\ra^2=1, \quad\la\chi\ra^2=\la\Psi_{+}\ra^2.
\label{av-square}\nn
\end{align}
For the gap ratio $\delta$ \Eref{Psi+} gives
\begin{equation}
\delta\equiv\frac{\Delta_1}{\Delta_2}
 =\frac{\Psi_{1+}}{\Psi_{2+}}
 =\frac{\chi_1}{\chi_2}
 =\mathrm{sign}(U)
   \sqrt{\frac{c_2}{c_1}}\sqrt{\frac{\epsilon+\xi}{\epsilon-\xi}}.
\end{equation}
Whence the moments of the anisotropy function read
\begin{equation}
\la\chi^n\ra=\frac{\la\Delta^n\ra}{\la\Delta^2\ra^{n/2}}
 =\frac{c_1\delta^n+c_2}{(c_1\delta^2+c_2)^{n/2}}.
\end{equation}
Using the general results formulated earlier, we find the transition
temperature of the clean two-band superconductor:
\begin{equation}
T_{c0}=\frac{2\gamma\bar{\epsilon}}{\pi}\mathrm{e}^{-\pi/\lambda_+}.
\label{2b-clean}
\end{equation}
Equation (\ref{T-line}) for the critical curve within this model can
be simplified to the form:
\begin{align}
\pi g\left[\eta +(c_1-c_2)\xi+2c_1c_2U - c_1 c_2f d \right] \nn\\
=(1-\lambda_+ f)(1- \lambda_- f).
\label{2b-T-line}
\end{align}
Here we have denoted $d=\mathrm{det}\,\hat{V}=W_1W_2-U^2,$ and
abbreviated the functions $f(\Tc,\tau)$ and $g(\Tc,\tau)$ as $f$ and
$g$, respectively, cf. Eqs.~(\ref{f}), (\ref{g}). It is convenient
to introduce the dimensionless scattering rate $x_0=(2\pi \kb
T_{c0}\tau/\hbar)^{-1}$ and the dimensionless transition temperature
$\theta=\Tc(\tau)/T_{c0}$; $x_c=x_0/\theta$. In terms of these
variables the functions $f$ and $g$ read:
\begin{align}
f&=\frac{1}{\lambda_+}-\frac{1}{\pi}\ln\theta-g,\nn\\
\pi g&=\digamma\left(\frac{x_0}{\theta}+\frac{1}{2}\right)
       -\digamma\left(\frac{1}{2}\right).
\label{fg}
\end{align}
The equation for \Tc\ finally reads
\begin{align}
\label{eq:Tc2band}
g\left[\la\chi\ra^2\lambda_{+}+\left(1-\la\chi\ra^2\right)\lambda_{-}-f\lambda_{+}\lambda_{-}\right]\nn\\
=\left(1-f\lambda_{+}\right)\left(1-f\lambda_{-}\right),
\end{align}
where
\begin{equation}
\la\chi\ra^2=\frac{\left(c_1\delta+c_2\right)^2}{c_1\delta^2+c_2},
\qquad c_2=1-c_1.
\end{equation}
To give a realization of the possible dependence $\theta (x_0)$, we
have made a numerical calculation setting\cite{Golubov:02}
$\delta=2.63,$ $\lambda_{+}=1.02,$ $\lambda_{-}=0.45,$ and
$c_1=0.422$; other authors give slightly different values, cf.
Refs.~\onlinecite{Wang:01,Bouquet:01,Drechsler:01,Choi:01,Papaconstantopulos:01}.
The results for $\theta(x_0)$ and $\Delta C/C_n(\Tc)$ vs $x_c$ are
shown in \Fref{fig:1}.

In the asymptotic regions of moderate and extreme dirt
\Eref{hohenberg} is valid with
$$
\kappa
=\left[\frac{1}{2}\left(1+\frac{(c_1-c_2)\xi}{\epsilon}\right)+\frac{c_1c_2U}{\epsilon}\right]^{-1}
=\frac{c_1\delta^2+c_2}{\left(c_1\delta+c_2\right)^2}
$$
for the clean and moderate dirt cases, and
$$
\bar{\kappa}=\frac{c_1^3W_1^2+c_2^3W_2^2+c_1c_2(2\eta+U)U}{
 \left(c_1^2W_1+c_2^2W_2 + 2c_1c_2U\right)^2}
$$
for the extreme dirt case. The order parameter in the moderate dirt
range is proportional to $\Psi_+$. In the range of strong disorder it
reads:
\begin{equation}
\chi=\left(\begin{array}{l}
c_1W_1+c_2U-\pi^{-1}c_1c_2d\,\ln(\bar{\epsilon}\tau/\hbar)\\
c_1U+c_2W_2-\pi^{-1}c_1c_2d\,\ln(\bar{\epsilon}\tau/\hbar)
\end{array}\right).
\label{2b-order}
\end{equation}
In the limit of extreme disorder
$\ln(\bar{\epsilon}\tau/\hbar)\ll W_{1,2}^{-1}$
it tends to the limiting value:
\begin{equation}
\chi_{\mathrm{extr}}=\bar{V}=\left({
c_1W_1+c_2U\atop
c_1U+c_2W_2}
\right).
\label{2b-extra}
\end{equation}
It is worthwhile to note that at $W_1=W_2=U$ the anisotropy
parameter $\kappa$ is equal to 1 for any value of the scattering
rate independently on the values $c_1,c_2$, and all
thermodynamical values do not depend on $\tau$ similarly to the
completely isotropic case.
%
%

\subsubsection{Specific heat discontinuity}

We note that the normalization factors of the gap-anisotropy function
and the superconducting gap cancel each other in the formulae for the
experimentally measurable jump of the specific heat. Therefore, we can
use the normalization $\la\chi^2\ra=1$ without loss of generality. The
calculations made above can be used directly to find the averages
necessary for the calculation of the specific heat. The influence of
the disorder on the relative specific heat discontinuity for a
two-band superconductor is shown in \Fref{fig:1}~(b). In the extreme
dirty limit we find that the relative specific heat discontinuity
tends to its isotropic value in agreement with the fact that the
density of states becomes isotropic in this limit.\cite{Pokrovsky:96}

\subsection{Dirty isotropic alloys}

It is straightforward to verify that in the isotropic case
($\chi=\mathrm{const}$ on the Fermi surface) the coefficients $a_0$
and $b$ do not depend on $\tau$. Thus, neither the energy gap, nor the
specific heat are influenced by impurities, in accordance with the
Anderson theorem.\cite{Anderson:59} Analyzing \Eref{T-line}, we
conclude that the transition temperature also does not depend on the
scattering rate. Indeed, in the isotropic case the eigenfunctions of
the operator $\hat{V}$ are spherical harmonics
$\Psi_{lm}=Y_{lm}(\theta,\varphi)$.  Among the latter only one,
$\Psi_{00},$ has a nonzero average. Thus, \Eref{T-line} takes the
simple form:
\begin{equation}
g(T_c,\tau)+f(T_c,\tau)=V_0^{-1}. \label{is-T-line}
\end{equation}
According to the definitions (\ref{f}) and (\ref{g}), the sum
$f(T,\tau)+g(T,\tau)=\ln\frac{\bar{\epsilon}}{2\pi \kb T}-\digamma
(1/2)$ does not depend on the scattering rate. Hence, $T_c$ also does
not depend on the scattering rate.
%

\subsection{Separable approximation}

\begin{figure}[t]
\centering
\includegraphics[width=0.8\columnwidth]{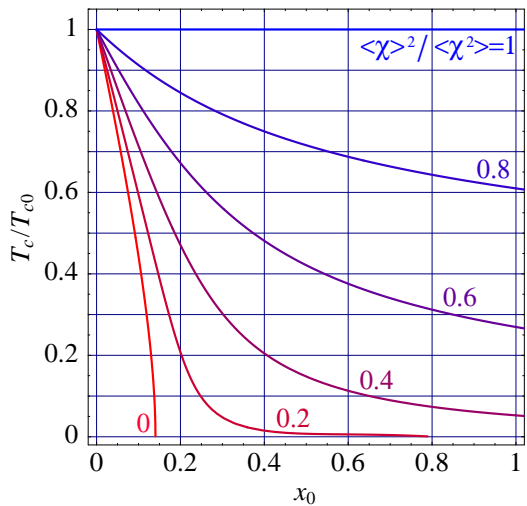}
\caption{Critical curves $\theta=T_c/T_{c0}$ vs $x_0=\hbar/2\pi\kb
T_{c0}\tau$ for different gap anisotropies
$\la\chi\ra^2/\la\chi^2\ra=$~0, 0.2, 0.4, 0.6, 0.8, and 1. For an
isotropic superconductor, $\la\chi\ra=1$, the critical temperature is
disorder independent while for $p$- or $d$-type superconductors (as
some CuO$_2$ superconductors are) $\la\chi\ra=0$ and the transition
line is exactly the same as the
Abrikosov-Gor'kov\protect\cite{AG,Abrikosov:88} curve for magnetic
impurities in isotropic superconductors. \label{fig:2} }
\end{figure}

The separable approximation
\begin{equation}
V(\mathbf{p},\mathbf{p}')=\sum_n V_n
\Psi_{n}^*(\mathbf{p})\Psi_{n}(\bp') \approx V_0 \chi_{\bp}\chi_{\bp'},
\end{equation}
where $\Psi_{0}(\mathbf{p})\equiv\chi_{\bp}$ is very often used for
modelling of the gap anisotropy in superconductors
$\Delta_{\bp}\approx\Xi(T)\chi_{\bp}$.  As we demonstrated earlier,
this approximation is valid in the range of clean and moderately dirty
superconductors. Applying this approximation to the equation for \Tc\
in two-band model \Eref{eq:Tc2band} we obtain Moskalenko and
Palistrant,\cite{Moskalenko:65} Abrikosov\cite{Abrikosov:93} and
Kogan's equation\cite{Kogan:02}
\begin{equation}
\ln \frac{T_{c0}}{T_c(\tau)}
=\left(1-\frac{\la\Delta_\mathbf{p}\ra^2}{\la\Delta_\mathbf{p}^2\ra}\right)
\left[\digamma\left(x_c+\frac{1}{2}\right)
-\digamma\left(\frac{1}{2}\right)\right],
\label{eq:AG}
\end{equation}
where\cite{Moskalenko:65}
$1/\tau=\frac{1}{2}(1/\tau_{12}+1/\tau_{21})$, and $1/\tau_{12}$ and
$1/\tau_{21}$ are rates of interband scattering.  The results of the
numerical solution of this equation are depicted on \Fref{fig:2}.  For
superconductors with zero averaged gap $\la\Delta_\mathbf{p}\ra=0$
which are $p$- and $d$-type superconductors, for example, this
equation formally coincides with the Abrikosov-Gor'kov
result\cite{AG,Abrikosov:88} for superconductors with magnetic
impurities; superconductivity disappears at the critical value
$x_0=1/4\gamma=0.1404.$ For weak disorder we have
\begin{equation}
T_{c0}-T_c\approx \frac{\la\chi^2\ra-\la\chi\ra^2}{\la\chi^2\ra}
\frac{\pi\hbar}{4\kb \tau}\ll T_{co}.
\end{equation}
In such a way one of the most important properties of multigap and
anisotropic superconductors is that the nonmagnetic impurities are
pair breaking, similar to magnetic impurities in conventional
superconductors.\cite{Pokrovsky:96,Moskalenko:65} A similar influence
of structural defects was discussed by Abrikosov\cite{Abrikosov:95}
for triplet superfluids. The reduction of the critical temperature by
disorder has been observed for layered
cuprates,\cite{Takabatake:88,Maple:97,Xiao:90,Moskalenko:91} for
impurity scattering in triplet superconductor\cite{Dalichaouch:95}
UPt$_3$ and recently for MgB$_2$.\cite{Buzea:01} Only dimensionless
ratios of the gap function moments, like $\la\chi\ra^2/\la\chi^2\ra$
in \Eref{eq:AG}, or $\la\chi^2\ra^2/\la\chi^4\ra$ in \Eref{eq:jump}
are relevant for the thermodynamics of superconductors. This explains
why strongly anisotropic-gap layered cuprates were seemingly
successfully analyzed as two-band superconductors\cite{Moskalenko:91}
(this reference is a comprehensive review of the properties of
multigap superconductors), and vice versa why the first prominent
two-gap superconductor MgB$_2$ could be analyzed as if it were a
single-band anisotropic-gap superconductor;\cite{Haas:01}
cf. \Rref{Mishonov:02b}.

%


\section{Discussion and Summary}

In order to investigate $\Delta C/C_n(\Tc)$ versus $1/\tau\Tc$
correlations it is necessary to have a good method for the
determination of $\tau.$ This could be far-infrared measurement of
the high-frequency conductivity, or merely the static resistivity.
For polycrystalline materials one has to use compounds like
MgB$_2$ with good connections between grains. The variation of the
residual resistivity can be produced by radiation.

For MgB$_2$ the gap ratio $\delta=\Delta_1/\Delta_2$ can be determined
by spectroscopic measurements, and the ratio of the DOS $\nu_1/\nu_2$
can be calculated from first principles. Under these conditions, for
high-quality clean samples we can evaluate the up-shift of the $\Delta
C/C_n(\Tc)$ curve, and the influence of strong-coupling
effects. Another complication is related to the variation of the
electron wave functions in the two bands. This can lead to different
scattering rates in the two bands. We consider that the scattering
rate in the $\sigma$-band with bigger gap is more important for the
disorder reduction of the specific heat jump. If the scattering time
cannot be determined by spectroscopic measurements we can use the
value of the resistivity at the critical temperature
$\rho_\mathrm{el}(\Tc)/\Tc$. Thus, the dependence of $T_c$ and the
relative specific heat jump on resistivity at $T=T_c$ will be given by
our formula and \Fref{fig:2}, \Fref{fig:1}~(b), with the scale of the
abscissa being a fitting parameter. We expect that the derived
weak-coupling formula can be as useful for the analysis of $\Delta C$
versus $\rho_\mathrm{el}$ correlations as the weak-coupling theory was
successful in describing the temperature dependence of the specific
heat for the clean MgB$_2$ samples.\cite{Bouquet:01} Thus, we conclude
that the weak-coupling theory of the impurity reduction of the
specific heat jump can reveal the main trend and qualitative
properties of the effect.

Irradiated superconductors are also a good example for the
application of the present theory. Furthermore, let us note that
conventional dirty superconducting alloys, for which a big enough
series of samples with continuously changing resistivity can be
prepared, are the best tools to investigate the influence of
disorder on the thermodynamics of superconductors.

Finally, let us summarize our results. (\textit{i}) In anisotropic
superconductors the transition temperature is suppressed by disorder
like $T_c\sim\tau^{\kappa-1},$ where
$\kappa=\la\chi^2\ra/\la\chi\ra^2$ is an anisotropy parameter which
is a slowly varying function of $\tau$. (\textit{ii}) The order
parameter retains its angular dependence until
$\bar{\epsilon}\tau\gg1,$ whereas the DOS becomes isotropic for
$\ln\bar{\epsilon}\tau\ll V_0^{-1}.$ The anisotropy of the order
parameter can be probed in tunneling experiments.  (\textit{iii})
The specific heat is suppressed by disorder just like $T_c$, with an
accuracy of a slowly varying factor. (\textit{iv}) The relative jump
of the specific heat $\Delta C/C_n(T_c)$ is smaller than its
isotropic value in the clean limit. It is enhanced by disorder
tending to its isotropic limit in the extreme disordered case.
(\textit{v}) In isotropic superconductors, and the two-band model
with $\Delta_1=\Delta_2$ and arbitrary $c_1/c_2,$ $\Tc$ and $\Delta
C$ do not depend on the scattering rate $1/\tau;$ in particular this
is the case for $W_1=W_2=U.$

\begin{acknowledgments}
This work was supported by the Flemish Government Programme GOA and
IUAP. VLP acknowledges support from NSF under the grant DMR0072115 and
by the Humboldt Foundation. He is also thankful to Prof.~W.~Selke and
RWTH Aachen for the hospitality.
\end{acknowledgments}

%


\begin{thebibliography}{99}
\bibitem{note} There is a misprint in Eq.~(81) of \Rref{Pokrovsky:96}:
the factor $x_c$ in the second term of the expression for the
coefficient $a$ was missed; $\pi^2$ in Eq.~(78) was misprinted as
$\pi^3$.
\end{thebibliography}
\end{document}